\documentstyle[twocolumn,floats,prl,aps,epsf,psfig,amsfonts,amssymb]{revtex}
\def\al{Alfv\'en}
\def\b{\textrm{b}}
\def\d{\textrm{d}}
\def\dt[#1]{\frac{\partial #1}{\partial \tau}}
\def\dz[#1]{\frac{\partial #1}{\partial z}}
\def\ddz[#1]{\frac{\partial^2 #1}{\partial z^2}}
\def\dvp[#1]{\frac{\partial #1}{\partial v_\|}}
\begin{document}
\title{Asymptotic Theory of Particle Trapping in Coherent Nonlinear Alfv\'en 
Waves}
\author{M.V. Medvedev \cite{mm}, P.H. Diamond \cite{pd}, 
M.N. Rosenbluth \cite{pd},}
\address{Physics Department, University of California, San Diego, La Jolla, 
CA 92039-0319}
\author{V.I. Shevchenko}
\address{ECE Department, University of California, San Diego, La Jolla, 
CA 92039-0407}
\maketitle
\draft
\begin{abstract}
A fully nonlinear, time-asymptotic theory of resonant particle
trapping in large-amplitude quasi-parallel Alfv\'en waves is presented.
The effect of trapped particles on the nonlinear dynamics of
quasi-stationary Alfv\'enic discontinuities and coherent Alfv\'en waves
is highly non-trivial and forces to a significant departure of
the theory from the conventional DNLS and KNLS equation models.
The virial theorem is used to determine the time-asymptotic
distribution function.
\end{abstract}

The magnetic fluctuations frequently observed in Solar Wind and 
Interstellar Medium plasma have been the subject of protracted and intense
observational and theoretical scrutiny. It is likely that these fluctuations
are nonlinear \al\ waves, in which the ponderomotive coupling of \al ic
magnetic field energy to ion-acoustic quasi-modes has modulated the phase 
velocity $v_A$, and so caused steepening and formation of discontinuities
\cite{us,before,us:PoP1}.
Such rotational and directional discontinuities have indeed been observed
in the Solar Wind, and are probably quasi-stationary waveform remnants of
nonlinearly evolved \al\ waves \cite{RD}.

Beginning with the work of Cohen and Kulsrud \cite{CohenKulsrud},
the theory of quasi-parallel, nonlinear \al\ waves has received a great
deal of attention \cite{DNLS} and has spawned in a variety of modifications 
of the wave envelope evolution equation, referred to as the Derivative 
Nonlinear Schr\"odinger (DNLS) equation. However, almost all attention has 
been concentrated on developing and extending the fluid theory of such 
waves, leaving issues of particle kinetics aside. Nevertheless, some 
attempts to incorporate particle dynamics into the DNLS model have been 
made, both analytically \cite{particleDNLS} and (very extensively) via
particle- and hybrid-code simulations \cite{simulations}. Progress in
constructing an analytical kinetic-MHD model of nonlinear
coherent \al\ waves occured recently by the self-consistent inclusion of
linear Landau damping \cite{before,us:PoP1} and gyro-kinetic (e.g., 
ion-cyclotron) effects \cite{us:PoP1}. However, even in these treatments, 
wave-particle resonant interaction is treated perturbatively
and calculated using the linear particle propagator. This technique
fails for a large-amplitude wave propagating in a finite-$\beta$
plasma (here $\beta$ is the ratio of kinetic and magnetic pressure)
because of non-perturbative effects associated with particle trapping
in the field of the wave. In this Letter, we extend the
theory of `kinetic' nonlinear \al\ waves to the strongly nonlinear
regime where trapped particles are important.

In the finite-$\beta$, isothermal regimes typical of the Solar Wind
(i.e., $c_s\sim v_A,\ T_e\sim T_i$) at 1~AU, resonant interaction of the 
plasma with ion-acoustic quasi-modes is a critical constituent of the
wave dynamics. The very existence of rotational discontinuities is due to 
the nonlinear coupling of \al\ waves to (linear) Landau dissipation 
\cite{us}. Here, linear Landau dissipation refers to damping calculated 
perturbatively, assuming a Maxwellian particle distribution function (PDF), 
and thus with a time-independent rate coefficient. This mechanism enters 
the \al\ wave dynamics {\em nonlinearly} (i.e., in proportional to the 
magnetic energy density of the wave train) because it enters 
a functional with
the parallel ponderomotive force $\propto\partial_z(\tilde B_\bot^2/8\pi)$. 
The `kinetic' wave equation, called the Kinetic Nonlinear Schr\"odinger 
(KNLS) equation, is \cite{before,us:PoP1}:
\begin{eqnarray}
\dt[\b]&+&v_A\dz[]\!\left(m_1\b|\b|^2+m_2\b\,\frac{1}{\pi}
\int_{-\infty}^\infty\frac{{\cal P}}{(z'-z)}|\b (z')|^2\d z'\right)
\nonumber\\
&+&i\frac{v_A^2}{2\Omega_c}\ddz[\b]=0 ,
\label{KNLS}
\end{eqnarray}
where $\b=(\tilde B_x+i\tilde B_y)/B_0$ is the normalized complex 
wave amplitude, $\Omega_c$ is the ion-cyclotron frequency,
the coefficients $m_1$ and $m_2$ are functions of $\beta$ and $T_e/T_i$ only 
(see \cite{us:PoP1}), and ${\cal P}$ means the principal value integration.

Obviously, particles which are near resonance with 
the wave ($v\simeq v_A$) will be trapped by the ponderomotive potential 
(or equivalently, by the electrostatic fields of driven ion-acoustic 
perturbations). Particle bounce motion significantly modifies the PDF near 
resonance, since trapped particle phase mixing results in flattening of 
the PDF (for resonant velocities) and formation of a {\em plateau}. 
Thus, the {\em linear} calculation of the Landau dissipation, 
while correct for times short compared to the typical  bounce (trapping) 
time, $\tau\ll\tau_{tr}$, fails for quasi-stationary waveforms for times 
$\tau\gtrsim\tau_{NL}\gg\tau_{tr}$ ($\tau_{NL}$ is the typical nonlinear 
wave profile evolution time). Hence, Landau dissipation should be
calculated {\em non-perturbatively} to determine the resonant particle 
response to the nonlinear wave.

Of course, the nonlinear Landau damping problem is, in general, not 
{\em analytically} tractable, as it requires explicit expressions for 
{\em all} particle trajectories as a function of initial position and time. 
Such trajectories cannot be explicitly calculated for a potential
of {\em arbitrary shape}. Usually, a
full particle simulation is required to obtain this information. 
In some cases, an approximate analytic expression
for the wave profile shape is known and may be assumed to persist, while the
wave amplitude varies. Calculations defined in this way has been implemented
for the special cases of sinusoidal \cite{ONeil} and solitonic 
\cite{NLDsolit} wave modulations. Other approaches either seek
the asymptotic ($\tau\to\infty$) PDF for a given (undamped) waveform
\cite{BGK}, or exploit the universality of the process of de-trapping 
of resonant particles from a wave potential of decreasing amplitude
\cite{mbi,comment}. These approaches, however, do not appear to be useful 
for the problem considered here.

The goal of this work is to investigate how trapped particles modify 
nonlinear wave evolution, assuming no restrictions on the shape of the
wave-packet modulation. Thus, the motion of particles is treated
self-consistently. We show that, in the two important limits of
short-time ($\tau\ll\tau_{tr}$) and long-time ($\tau\gg\tau_{tr}$) 
evolution, the problem admits {\em analytic} solutions. In the limit
$\tau\ll\tau_{tr}$, we recover conventional linear Landau damping.
This supports the validity of the KNLS theory as a means for studying 
the {\em emergence} of \al ic discontinuities. In the opposite limit 
$\tau\gg\tau_{tr}$, the {\em virial theorem} is used for determination of 
the time-asymptotic trapped particle response. 
Although the damping rate vanishes due to phase mixing, the effects 
of trapped particles are highly non-trivial, leading to a significant
departure of the theory from the familiar form of the DNLS and KNLS models. 
First, the power of the KNLS nonlinearity associated with resonant 
particles increases to {\em fourth} order when trapped particles are
accounted for. Second, the effective coupling 
now is proportional to the {\em curvature} of the PDF
at resonant velocity, $f_0''(v_A)$, and not its slope, $f_0'(v_A)$, as 
in linear theory. Third, the {\em phase density} of trapped particles
is controlled by the plasma $\beta$. Finally, we combine these to obtain 
the wave evolution equation which governs the {\em long-time} dynamics of 
quasi-stationary \al ic discontinuities. The equation is the principal
result of this Letter.

We should state here that particle trapping may be absent in
higher than one dimension. Indeed, for $k_\bot\rho_i\gg1$ ($k_\bot$ is
the perpendicular component of the wave vector and $\rho_i$ is the
ion Larmor radius), then the longitudinal \v Cerenkov resonance 
$\omega=k_\|v_\|$ is satisfied for all particles having $v=v_A/\cos\Theta$, 
but with $v_\bot$ arbitrarily large. Thus, all particles with velocities
$v\gtrsim v_A$ interact with a wave and a plateau cannot form, while a 
non-thermal tail of energetic particles may result instead.
However, if the magnitude of the ambient magnetic field is strong enough so
that $k_\bot\rho_i\ll1$, quasi-one-dimensionality is recovered. This last
situation is, in fact, typical for waves propagating in the Solar Wind.

For reasons of notational economy, let's introduce the trapping potential 
$U(z)\equiv\tilde B_\bot^2/8\pi n_0$, where $n_0$ is the unperturbed
plasma particle density. Then, the characteristic bounce frequency
\cite{ONeil} in our case is $\tau^{-1}_{tr}\simeq k\sqrt{U/m_i}$
($m_i$ is the ion mass). The characteristic nonlinear frequency
at which the wave profile changes appreciably is readily estimated
from Eq.\ (\ref{KNLS}) to be 
$\tau^{-1}_{NL}\simeq m_1 kv_A (\tilde B_\bot^2/B_0^2)$. From comparison of
these two time-scales, we conclude that the wave potential, as seen by
a trapped particle, is {\em steady-state} (i.e., roughly constant on
the particle bounce time) when $\tau_{NL}\gg\tau_{tr}$, so that
\begin{equation}
\tilde B_\bot/B_0\lesssim m_1^{-1}\sim1 .
\label{condition}
\end{equation}
That is, particle phase mixing is very {\em efficient} for weakly 
nonlinear waves. Note, this condition (\ref{condition}) is consistent 
with the derivation of the KNLS, for which $\tilde B_\bot/B_0\ll1$ 
(weak nonlinearity) is assumed.
Let's now rewrite Eq.\ (\ref{KNLS}) in a generic form:
\begin{equation}
\dt[\b]+v_A\dz[]\biggl(m_1\b\;\delta n_{NR}
+m_2\b\;\delta n_{R}\biggr)+i\frac{v_A^2}{2\Omega_c}\ddz[\b]=0 .
\label{GKNLS}
\end{equation}
Here $\delta n_{NR}$ is the density perturbation due to the 
{\em non-resonant} (bulk) response of the PDF. It is roughly
proportional to $|\b|^2$. 
$\delta n_R$ is the {\em resonant} particle contribution. It is responsible 
for {\em strongly nonlinear} feedback via the distortion of the PDF by 
a wave. It was also responsible for linear damping in the KNLS equation.
It is interesting that the very possibility to write the generalized KNLS
equation in the form (\ref{GKNLS}) relies on the intrinsic {\em time
reversibility} of the Vlasov equation, linear or nonlinear. Indeed,
one can formally write the resonant particle response as
\begin{equation}
\delta n_R\propto\chi_\|\hat{\cal K}\left[U(z)\right] ,
\label{nR}
\end{equation}
where $\hat{\cal K}$ is some normalized kinetic operator acting on a wave 
field. Time reversibility implies $\hat{\cal K}\hat{\cal K}=-1$, see 
\cite{us:PoP1}. This fact has been crucial for the derivation of KNLS.
The constant $\chi_\|$ plays a role of effective dissipation 
coefficient (thermal conductivity) in the linear Landau damping theory.

\begin{figure}
\psfig{file=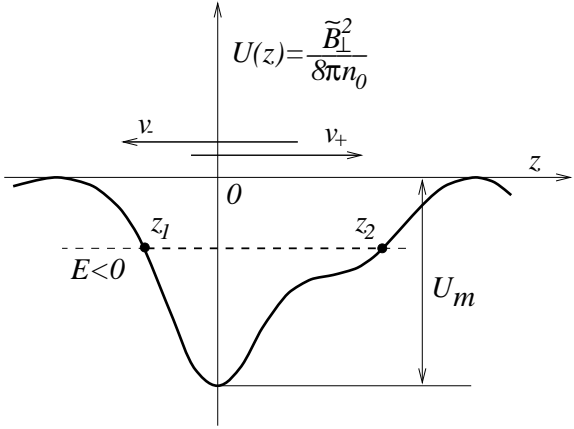,width=3in,angle=-90}
\caption{Trapping potential.}
\label{fig:potential}
\end{figure}

The resonant particle response is calculated using {\bf Liouville's theorem},
which states that {\em the local PDF is constant along particle 
trajectories}:
\begin{equation}
f(v,z,t)=f_0\left(v_\pm\left(E,z_0^\pm\right)\right) ,
\end{equation}
where $z_0^\pm=z_0^\pm\left(z,t,E;U(z)\right)$ is the 
{\em initial} coordinate of a particle of total energy $E$ which at time 
$t$ is at the point $z$ and has a velocity $v_\pm(E,z)$. 
Thus, $z_0^\pm$ is a solution of
${t=(\pm)\int_{z_0^\pm}^z
\left[\frac{2}{m_i}\left(E-U(z)\right)\right]^{-1/2}\d z}$\ . 
By definition 
$\delta n_R=\int_{\Delta v_{res}}\d v\left(f-f_0^{(t=0)}\right)$:
\begin{equation}
\delta n_R=\frac{1}{\sqrt{2m_i}}\sum_{(\pm)}\int_{\Delta E_{res}}
\!\!\!\!\!\!
\frac{f_0\left(v_\pm\left(E,z_0^\pm\right)\right)%
-f_0\left(v_\pm\left(E,z\right)\right)}%
{\sqrt{E-U(z)}}\; \d E\ .
\label{dnR}
\end{equation}
Here the sum is over particles moving to the right (+) and to the left (-),
as in Fig.\ \ref{fig:potential}. The integration is over the resonant
(negative) energies of trapped particles, $U_m\le E \le 0$ with $U_m$
being the amplitude of the potential.

Let's first consider the short-time limit, $\tau\to 0$.
Then the following approximations are valid. First (i), the effective 
resonance width \cite{ONeil} is 
$\Delta E_{res}=(\Delta v_{res})^2\sim 1/(k\tau)^2\to\infty$, 
as $\tau\to0$. Second (ii), the particle velocity change is negligible
$|v(z,E)-v(z_0,E)|\ll\sqrt{U(z)}$, so that (iii), the particle position
is roughly proportional to time $z_0^\pm\simeq z\pm \tau\sqrt{2E/m_i}$.
Finally (iv),
the PDF response can be linearized (in the wave frame moving with $v_A$) as
$f_0(v)\simeq f_0(v_{res})+v\,f_0'(v_{res})$. Then Eq.\ (\ref{dnR})
may be estimated as 
\begin{eqnarray}
\biggl.\delta n_R\biggr|_{t\to 0}
&\simeq&\frac{f_0'(v_A)}{2m_i}\sum_{(\pm)}\ \pm\!\!
\int_{U_m}^{\pm\infty} \frac{\d E}{E}\ U\!\left(z\pm\tau\sqrt{2E/m_i}\right)
\nonumber\\
&\simeq&\frac{\pi f'_0(v_A)}{m_i}\left\{\frac{1}{\pi} 
\int_{-\infty}^\infty\frac{\cal P}{(z'-z)}U(z')\d z'\right\} .
\end{eqnarray}
Here, we first used  (i) to extend the integration over $\Delta E_{res}$ to
$\pm\infty$, and then (iii) and (i) to expand the denominator in (\ref{dnR})
for $E\gg U$. Finally, we took the $\tau\to 0$ limit.
Compared to Eq.\ (\ref{nR}), the particle operator $\hat{\cal K}$
is replaced by the Hilbert operator, $\hat{\cal H}$, given by the 
expression in curly brackets. It is nonlocal and satisfies the 
time-reversibility condition $\hat{\cal H}\hat{\cal H}=-1$. 
The effective dissipation coefficient is simply $\chi_\|=\pi f'_0(v_A)/m_i$. 
Thus the KNLS equation (\ref{KNLS}) is recovered \cite{us:PoP1}.

To treat the $\tau\to\infty$ limit, we recall that for the times
$\tau\gtrsim\tau_{NL}$ steady-state waveforms (discontinuities) have
formed. Thus, particles are trapped in these adiabatically
changing potentials. Hence, we may employ the {\bf virial theorem}, which
states that {\em for any finite motion} in a potential
~$\tilde U(z)=U(z)-U_m$~ [i.e., ${\tilde U(z)\ge0}$] {\em
the (period) averaged \underline{kinetic} and \underline{potential} energies 
are related by}
\begin{equation}
2\langle K(z)\rangle=n\langle\tilde U(z)\rangle .
\label{virial}
\end{equation}
Here $\tilde U(z)$ is a {\em homogeneous} function of its argument 
of order $n$, i.e., $\tilde U(az)=a^n\tilde U(z)$.
The resonance width is easily estimated to be
$\Delta v_{tr}\simeq\sqrt{2|U_m|/m_i}$ with 
$|U_m|\sim\tilde B_\bot^2/8\pi n_0$. Thus, for weak nonlinearity, 
the resonance is {\em narrow}:
\begin{equation}
\frac{\Delta v_{tr}}{v_A}\sim\frac{\tilde B_\bot}{B_0}\ll1 .
\label{width}
\end{equation}
Hence, an expansion of the PDF is valid, so that (in the wave frame):\
$f_0(v_\pm)\simeq f_0(v_A)
\pm v_\pm f_0'(v_A)+\left(v_\pm^2/2\right)f_0''(v_A)$. With this in hand
and using Eq.\ (\ref{virial}) and $\langle U\rangle+\langle K\rangle=E$,
we calculate the resonant particle contribution, Eq.\ (\ref{dnR}):
\begin{eqnarray}
& &\biggl.\langle\delta n_R\rangle\biggr|_{\tau\to\infty}
\simeq f_0''(v_A)
\sqrt{\frac{2}{m^3_i}}\sqrt{|U(z)|}
\nonumber\\ & &\quad\times
\left[\frac{n}{n+2}\left(|U_m|-|U(z)|\right)-\frac{2}{3(n+2)}|U(z)|\right] .
\label{large-t}
\end{eqnarray}
Note that the term $\propto f_0'(v_A)\,\left[v(z_0^+)-v(z_0^-)\right]$ 
vanishes identically because 
$\langle U(z_0^+)\rangle=\langle U(z_0^-)\rangle$.
Thus damping is absent. Since 
$\langle{\cal K}\rangle\langle{\cal K}\rangle\not={\cal K}{\cal K}=-1$,
we can, however, only {\em estimate} [from Eq.\ (\ref{nR})] the coupling 
constant to be $\chi_\|\sim f_0''(v_A)\sqrt{2/m^3_i}$.
The index $n$ is formally not defined for an arbitrary potential. One
may, however, {\em estimate} it comparing the calculated bounce period
in the homogeneous potential and ``actual'' one determined numerically
for a known $U$, i.e., 
\begin{mathletters}
\begin{eqnarray}
& &T_{hom}(E)=\left|E\right|^{\frac{1}{n}-\frac{1}{2}} ,\\
& &T_{act}(E)=\sqrt{\frac{m_i}{2}}
\int_{z_1}^{z_2}\frac{\d z'}{\sqrt{E-U(z')}} .
\end{eqnarray}
\end{mathletters}
It is interesting that the limit $n\to\infty$ encompasses two frequently
encountered shapes of a wave packet, namely the {\em solitonic} and 
{\em rectangular} (i.e., deep narrow well) forms. In fact, for these cases
as well as for any rather {\em anharmonic} potentials ($n\gg2$) the 
resonant particle response (\ref{large-t}) is independent of $n$ and
takes on a very simple form:
\begin{equation}
\biggl.\langle\delta n_R\rangle\biggr|_{{\tau\to\infty}\atop{n\to\infty}}
\simeq f_0''(v_A)\sqrt{\frac{2}{m^3_i}}\sqrt{|U(z)|}
\left(|U_m|-|U(z)|\right) .
\end{equation}
Thus, in the long-time limit, $\tau\gg\tau_{tr}$, the damping rate vanishes
due to phase mixing. Nevertheless, the resonant particles still contribute
the wave dynamics, in that 
\begin{equation}
\langle\delta n_R\rangle\sim f_0''(v_A) |\b|^3 ,
\end{equation}
thus determining a new nonlinear wave equation.

\begin{figure}
\psfig{file=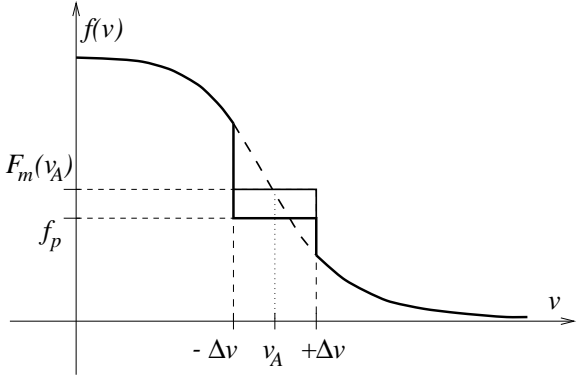,width=3in,angle=-90}
\caption{Asymptotic, $\tau\to\infty$, particle distribution function.}
\label{BGK}
\end{figure}

To estimate the number of trapped particles, we use a BGK-type 
(Bernstein-Green-Kruskal) approach
\cite{BGK}. This allows us to find the PDF such that the wave
of a given profile is not dissipated by the Landau mechanism. 
In our problem there is only one resonance at $v\simeq v_A$, since 
all modes are coherent. At large times, particle bounces result in 
flattening of the PDF at resonance so that $f_0'(v_A)\to 0$.
The height of this plateau (i.e., the phase density
of trapped particles) depends on the wave evolution at earlier times.
We take the ``unperturbed'' (trial) PDF as superimposed plateau and 
Maxwellian, as in Fig.\ \ref{BGK}:
\begin{equation}
f_0(v)=F_m(v)+\left[f_p-F_m(v)\right]\,\Theta_{v_A}\left(\Delta v\right) ,
\end{equation}
where $F_m(v)$ is Maxwellian, the $\Theta$-function is defined as
$$
\Theta_{v_A}\left(\Delta v\right)=\left\{
\begin{array}{ll}
1, & \textrm{if $(v_A-\Delta v)\le v\le (v_A+\Delta v)$;}\\
0, & \textrm{otherwize;}
\end{array}\right. 
$$
and $f_p$ is the constant to be determined. The coefficient $f_p$ has a
simple meaning of the phase density of trapped particles after the 
plateau has been formed. Thus, the state with $f_p>F_m(v_A)$ corresponds to 
a {\em clump} on the PDF and that with $f_p<F_m(v_A)$ corresponds 
to a {\em hole}.
The kinetic equation for a perturbation of the PDF is:
\begin{equation}
\left(-i\omega_k+\gamma_k+ik_\|v_\|\right)\tilde f_{\omega,k}
=ik_\|U_{k_\|}\dvp[f_0(v)]\ .
\end{equation}
By definition, ${\delta n_{\omega,k}=\int f_{\omega,k}\d v}$. Then, for
$\gamma_k\ll\omega_k=k_\|v_A$ and $\Delta v/v_A\ll1$, we obtain:
\begin{eqnarray}
\delta n_R&=&\sum_k e^{ik_\|z}U_{k_\|}~2k_\|^2\Delta v 
\nonumber\\ & &\quad\times
\frac{-\left(i\gamma_k/k_\|\right)F'_m(v_A)+\left[F_m(v_A)-f_p\right]}%
{\gamma_k^2+k_\|^2\Delta v^2}\ .
\end{eqnarray}
Looking for the stationary solution, $\gamma_k=0$, of the general KNLS
equation (\ref{GKNLS}) and neglecting dispersion, we have
$\partial_z[\b\delta n_{NR}+\b\delta n_R]=0$. Consequently,
$$
m_1\b|\b|^2+m_2\b\sum_k e^{ik_\|z}|\b|^2_{k_\|}
\frac{F_m(v_A)-f_p}{\Delta v/2}=0\ .
$$
We thus obtain the trapped particle phase density:
\begin{equation}
f_p=F_m(v_A)+\frac{m_1}{m_2}\frac{\Delta v}{2v_A^2}\ ,
\end{equation}
with $\Delta v\equiv\Delta v_{tr}\simeq v_A(\tilde B_\bot/B_0)$.
$F_m(v_A)$ is the particle phase density in the absence of trapping.
Recalling that $m_1$ and $m_2$ are functions of $\beta$ and 
$\chi_\|\propto f_0'(v_A)$ \cite{us:PoP1}, we conclude that
there must be an {\em under-population} of trapped particles [$f_0<F_m(v_A)$]
in a low-$\beta$ plasma ($\beta\lesssim1$) and an {\em over-population}
[$f_0>F_m(v_A)$] in a high-$\beta$ plasma ($\beta\gtrsim1$).

Finally, consider the there is weak wave damping not associated with 
\v Cerenkov
resonance (e.g., as in ion-cyclotron or collisional damping). Then the wave 
amplitude will slowly decrease, keeping resonant particles trapped. The 
following adiabatic invariant is thus conserved:
\begin{equation}
J=\oint p_\|\d z\simeq const ,
\end{equation}
i.e., $\langle |v_\| |\rangle(z_2-z_1)\simeq const$. From Eq.\ (\ref{KNLS}),
one can estimate $\Delta z\sim(\Omega_c/v_A)(\tilde B_\bot/B_0)^{-2}$.
Hence, $\Delta v_\|\sim(\tilde B_\bot/B_0)^{2}$.
The resonance width is, however, $\Delta v_{tr}\sim(\tilde B_\bot/B_0)$.
Thus, 
\begin{equation}
\frac{\Delta v_\|}{\Delta v_{tr}}\sim\frac{\tilde B_\bot}{B_0} ,
\end{equation}
that is, the trapped particles will {\em condense} near the bottom of the 
potential well, as the wave amplitude decreases. 
This results in a decrease in the effective index $n$, which 
approaches the asymptotic limit $n\to2$. The BGK analysis given above is, 
however, then no longer applicable. It should be emphasized that trapped
particles condense in the bottom of the potential, rather than de-trap 
from it, as naively suggested in Ref.\ \cite{mbi}. Thus, no asymptotic, 
power-law damping exists in this case. Obviously, our considerations above
are rather generic and valid for a wide class of nonlinear wave systems
with quadratic nonlinearity and higher, and thus call the 
validity of the results of Ref.\ \cite{mbi} into general question.

To conclude, we have shown that the effects of the nonlinear PDF 
modification by a high-amplitude \al\ wave significantly modify the 
dynamics of such a wave. 
Even when phase mixing is efficient enough to quench linear Landau
dissipation, trapped particles produce finite a response which modifies 
the wave nonlinearity. The equation which explicitly describes the evolution 
of {\em quasi-stationary} \al ic discontinuities and asymptotic 
($\tau\to\infty$) dynamics of nonlinear 
\al\ waves, Eqs.\ (\ref{GKNLS},~\ref{large-t}), has been obtained.
this result constitutes the extension of the well established DNLS-KNLS
theory of quasi-parallel nonlinear \al\ waves to the strongly
nonlinear regime of particle trapping.
The phase density of trapped particles has been shown to be controlled
by the value of plasma $\beta$, as well as wave amplitude.

We would like to thank R.Z. Sagdeev for valuable
and interesting discussions. This work was supported by DoE grant
DE-FG03-88ER53275.


\begin{references}
\bibitem{comment} We want to comment here that the theory of 
Ref.\ \cite{mbi} is, probably, not as general as it is claimed, and 
the problem considered in this Letter is, in fact, one of the 
counter-examples; see end of this Letter.
\end{references}
\end{document}